\documentclass[aps,prl,twocolumn,superscriptaddress,nofootinbib,notitlepage,longbibliography]{revtex4-1}
\usepackage{bm}
\usepackage{hhline}
\usepackage{amsmath}
\usepackage{amssymb}
\usepackage{mathdots}
\usepackage{tabularx}
\usepackage{graphicx}
\usepackage{hyperref}
\usepackage{bbding}
\usepackage{tikz}
\usepackage{bbding} 
\usepackage{braket}
\usepackage{bm}

\begin{document}
 \title{Orbital Longitudinal Magneto-electric Coupling in Multilayer Graphene}
	\author{Jin-Xin Hu}
	\affiliation{Division of Physics and Applied Physics, School of Physical and Mathematical Sciences, Nanyang Technological University, Singapore 637371} 
    \affiliation{Department of Physics, Hong Kong University of Science and Technology, Clear Water Bay, Hong Kong, China}
      \author{Justin C. W. Song}\thanks{justinsong@ntu.edu.sg}
	\affiliation{Division of Physics and Applied Physics, School of Physical and Mathematical Sciences, Nanyang Technological University, Singapore 637371}

	\begin{abstract}
Magneto-electric coupling enables the manipulation of magnetization by electric fields and vice versa. While typically found in heavy element materials with large spin-orbit coupling, recent experiments on rhombohedral-stacked pentalayer graphene (RPG) have demonstrated a {\it longitudinal magneto-electric coupling} (LMC) without spin-orbit coupling. Here we present a microscopic theory of LMC in multilayer graphene and identify how it is controlled by a ``layer-space'' quantum geometry and interaction-driven valley polarization. Strikingly, we find that the interplay between valley-polarized order and LMC produces a butterfly shaped magnetic hysteresis controlled by out-of-plane electric field: a signature of LMC and a multiferroic valley order. Furthermore, we identify a nonlinear  LMC in multilayer graphene under time-reversal symmetry, while the absence of centrosymmetry enables the generation of a second-order nonlinear electric dipole moment in response to an out-of-plane magnetic field. Our theoretical framework provides a quantitative understanding of LMC, as well as the emergent magneto-electric properties of multilayer graphene.

	\end{abstract}
	\maketitle

\emph{Introduction.}---Multilayer graphene 
has garnered increasing attention and emerged as a versatile platform for van der Waals heterostructures~\cite{hass2008growth,nilsson2008electronic,zhang2009direct,song2018electron,xiong2024antiscreening}. A variety of exotic phenomena have been observed in multilayer graphene, including interaction-driven Chern insulators~\cite{han2024large,han2024correlated,sha2024observation,lu2024fractional,waters2024interplay,lu2025extended,dong2024theory,zhou2024fractional,dong2024anomalous}, ferromagnetism~\cite{zhou2022isospin,zhou2021half,dong2023collective,zhou2024layer,chen2023gate}, and superconductivity~\cite{zhou2021superconductivity,zhang2023enhanced,han2024signatures,li2024tunable}. Among these, rhombohedral-stacked pentalayer graphene (RPG) has recently been reported to exhibit multiferroicity~\cite{han2023orbital} manifesting a magnetic ordering that can switched by gate electric field. This phenomena is particularly striking since spin-orbit coupling in RPG vanishes. Instead, such multiferroicity arises hand-in-hand with an orbital magneto-electric coupling that mediates the transduction between electric and magnetic fields within materials. This transduction enables the control of magnetization (or polarization) through electric (or magnetic) fields~\cite{eerenstein2006multiferroic,dong2015multiferroic,fiebig2016evolution,spaldin2010multiferroics,vopson2015fundamentals,
liu2020magnetoelectric,huang2018electrical,huang2018prediction,ahn2022theory,xiao2022intrinsic,abouelkomsan2024multiferroicity}.

In the modern theory of magneto-electric coupling, major attention has focused on axion electrodynamics~\cite{li2010dynamical,sekine2021axion}, where axion magneto-electric coupling is quantized by the fundamental constant $e^2/h$~\cite{wang2015quantized,dziom2017observation,
morimoto2015topological,zhang2019topological,essin2009magnetoelectric,liu2020robust,lei2024capacitive,mei2024electrically,he2022topological}. Besides the topological magneto-electric effects, the spin-orbit coupling can also lead to the linear magneto-electric coupling in three-dimensional materials~\cite{scaramucci2012linear}. In contrast to the above seminal works, the emergent magneto-electric coupling in 
2D layered quantum materials found recently in the experiment of Ref.~\cite{han2023orbital} can persist even in the absence of spin-orbit coupling~\cite{das2024superpolarized}.

\begin{figure}
		\centering
		\includegraphics[width=1.0\linewidth]{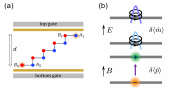}
		\caption{(a) Schematic illustration of the rhombohedral pentalayer graphene  embedded in a dual-gate device. With the aid of top and bottom gates, both the carrier density and out-of-plane electric field can be tuned. The distance between the top and bottom layers is $d$. (b) The layer-polarized orbital magnetic moment acquires a correction in response to $E$ field, while the layer electric dipole moment receives a correction in response to the $B$ field. Both of them contribute to an emergent orbital {\it magneto-electric} moment described in Eq.~\eqref{eq:EBcouple}.}
		\label{fig:fig1}
\end{figure}

Here we argue that the longitudinal magneto-electric coupling (LMC) in 2D layered materials is mediated by a layer-dependent quantum geometry manifesting as an out-of-plane control of magnetization by a gate electric field, and vice versa. In particular, we unveil a microscopic orbital {\it magneto-electric moment} that emerges from examining how the magnetic moment changes with an applied out-of-plane electric field as well as how layer dipole moment changes with an applied magnetic field (see Fig.~\ref{fig:fig1}). Critically, as we discuss below, valley polarization -- a unique correlated order in multilayer graphene -- plays a central role in the linear LMC observed in RPG by lifting the valley degeneracy of its electronic bands. Notably, we find that LMC can intertwine with the valley-polarized order yielding an ``electric field-valley polarization'' coupling. As we explain below this naturally produces a butterfly-shaped hysteresis recently observed in experiments~\cite{han2023orbital}.  

When valley polarization is turned off, we find layer mediated LMC can persist as 
nonlinear LMC. For example, such nonlinear LMC allows magnetic field to control layer polarization for noncentrosymmetric but time-reversal invariant systems. As an intrinsic effect, nonlinear LMC generally exists not only in multilayer graphene, but also in other van der Waals layered materials and includes contributions from both the Fermi-surface and the Fermi-sea terms. These highlight how the rich orbital magneto-electric effects in layered materials is mediated by the layer degree of freedom.

\vspace{2mm}
\emph{Orbital magneto-electric moment \& linear LMC in multilayers.}---Microscopically, the LMC can be generally described via:
\begin{align}
\label{eq:me_eq}\delta M = \chi_{me} E+\gamma_e E^2+... \\
\label{eq:em_eq}\delta P = \chi_{em} B+\gamma_b B^2+...
\end{align}
where $\delta M$ is the induced magnetization, $\delta P$ is the polarization change, and $E$ and $B$ are the applied electric and magnetic fields respectively; in what follows we will focus on the case where all fields, polarization, and magnetization are out-of-plane.
Here the linear magneto-electric coefficients are denoted $\chi_{me}$ and $\chi_{em}$ respectively;  a maxwell relations ensures $\chi_{em}=\chi_{me}$. Importantly, in order for nonzero $\chi_{me}$ and $\chi_{em}$ to manifest, time reversal and inversion symmetries need to be broken. In contrast, $\gamma_e$ survives centrosymmetry and $\gamma_b$ survives time reversal. 

To evaluate the longitudinal electric and magnetic responses in layered quantum materials, we adopt a semiclassical approach that incorporates
geometric phases and orbital magnetization. We begin by analyzing an effective field-dependent free energy $F(E,B)$ as~\cite{xiao2010berry,xiao2005berry}
\begin{equation}
\label{eq:free_energy}
F(E,B)=-\frac{1}{\beta}\sum_{n\bm{k}}\big\{1+\frac{e}{\hbar}B\tilde{\Omega}_{n\bm{k}}\big\}\phi\big[\varepsilon_{n\bm{k}} (E,B)\big],
\end{equation}
where $n,\bm{k}$ are the band index and wavevector, $\phi(\varepsilon)=\mathrm{ln}[1+e^{-\beta(\varepsilon-\mu)}]$. We model the field dependent Bloch band electronic energy via $\varepsilon_{n\bm{k}} (E,B)=\varepsilon_{n\bm{k}}^{(0)}+\delta \varepsilon_{n\bm{k}} (E,B)$. Here $\varepsilon_{n\bm{k}}^{(0)}$ is the bare Bloch band electronic energy without applied $E$ or $B$ fields, $\tilde{\Omega}_{n\bm{k}} $ is the Bloch band Berry curvature, $\beta = 1/k_B T$ is an inverse temperature, and $\mu$ is the chemical potential. $\delta \varepsilon_{n\bm{k}} (E,B)$ are field-induced corrections discussed in detail below. In the following, spin acts as a spectator degree of freedom since graphene systems possess negligible spin-orbit coupling.

As we now argue, the layer degree of freedom mediates an emergent orbital magneto-electric coupling in multilayer graphene. This can be revealed by examining how the Bloch electronic states change under an applied out-of-plane electric field $E$ and magnetic field $B$ via 
\begin{equation}
H(E,B)=H_0(\bm{k}) +u\hat{\bm{p}} -\hat{\bm{m}} B, \quad u = edE 
\end{equation}
where $u$ is the layer potential drop and $d$ is the distance between top and bottom layers [see Fig.~\ref{fig:fig1}(a)]. Here $H_0(\bm{k})$ is the bare material Hamiltonian such that $H_0(\bm{k}) |n \bm{k}\rangle = \varepsilon_{n\bm{k}}^{(0)} |n \bm{k}\rangle$ with $|n \bm{k}\rangle$ the Bloch eigenstate, $\hat{\bm{p}}$ is the layer-polarization operator (for example, $\hat{\bm{p}}=\sigma_z/2$ in the layer space for a bilayer system~\cite{gao2020tunable,zheng2024interlayer,fan2024intrinsic,hu2024colossal}) and $\hat{\bm{m}} = e(\hat{\bm{v}} \times \hat{\bm{r}} - \hat{\bm{r}}\times \hat{\bm{v}})/4$ is the orbital magnetic momentum operator~\cite{song2019low} that describes an orbital-Zeeman coupling with $\hbar\hat{\bm{v}}=\nabla_{\bm{k}}H_0(\bm{k})$ the velocity operator and $\hat{\bm{r}}$ the position operator.

By directly expanding the energy of the Bloch states $|n \bm{k} \rangle$ to order $\mathcal{O} (EB)$ we find the field-corrected band electronic energy reads as 
\begin{equation}
    \delta \varepsilon_{n\bm{k}}(E,B) = edp_{n\bm{k}}E - m_{n\bm{k}}B + \alpha_{n\bm{k}} EB 
    \label{eq:energy}
\end{equation}
where $p_{n\bm{k}}=\langle n|\hat{\bm{p}}|n\rangle$ is the layer dipole moment for band $n$ and 
$m_{n\bm{k}} = (e/2)\sum_{m\neq n}\bm{v}_{nm}\times \bm{r}_{mn}$ is the intraband orbital moment with $\bm{v}_{nm}$ and $\bm{r}_{nm}$ the interband velocity matrix elements and interband Berry connection respectively. 

While the first two terms in Eq.~(\ref{eq:energy}) represent the Stark shift and orbital Zeeman effect arising directly from the first-order expectation value, the last term in Eq.~(\ref{eq:energy}) $\alpha_{n\bm{k}}$ describes an orbital {\it magneto-electric moment}, emerging from a layer-dependent quantum geometry. To see this, notice that $u \hat{\bm{p}}$ perturbs the Bloch eigenstates $|n\bm{k}\rangle$. To first order in electric fields, the velocity matrix elements change as:  $\tilde{\bm{v}}_{nm} \to \bm{v}_{nm} + u\delta_u \bm{v}_{nm}$, where
\begin{equation}
   \delta_u \bm{v}_{nm}= \sum_{q\neq n}p_{nq}\bm{v}_{qm}/\varepsilon_{nq}^{(0)}+\sum_{l\neq m}p_{lm}\bm{v}_{nl}/\varepsilon_{ml}^{(0)}, 
   \label{eq:par_vnm}
\end{equation}
where $\varepsilon_{ln}^{(0)}=\varepsilon_{l\bm{k}}^{(0)}-\varepsilon_{n\bm{k}}^{(0)}$. In the same fashion, the changes to the interband Berry connections can be computed from the continuity equation $\bm{r}_{mn}^a=\langle m|i\partial_a|n\rangle = -i\hbar \bm{v}_{mn}^a/\varepsilon_{mn}$ (for $m\neq n$), see {\bf SI}~\cite{NoteX}. Critically, $\delta_u$ produces layer potential induced changes to the orbital magnetic moment and its Berry curvature: $\tilde{m}_{n \bm{k}} \to m_{n \bm{k}} + u\delta_u m_{n \bm{k}}$ and $\tilde{\Omega}_{n \bm{k}} \to \Omega_{n \bm{k}} + u\delta_u \Omega_{n \bm{k}}$ with 
\begin{equation}
\label{eq:par_mnk}
\begin{split}
  \delta_u m_{n\bm{k}}=-\sum_{m\neq n}\frac{e\hbar}{[\varepsilon_{nm}^{(0)}]^2}&\mathrm{Im}[\varepsilon_{nm}^{(0)}\delta_u\bm{v}_{nm}\times \bm{v}_{mn}+\\
  & v_{nm}^x v_{mn}^y (p_m-p_n)],
\end{split}    
\end{equation}
and
\begin{equation}
\label{eq:par_onk}
\begin{split}
  \delta_u \Omega_{n\bm{k}}=-\sum_{m\neq n}\frac{2\hbar^2}{[\varepsilon_{nm}^{(0)}]^3}&\mathrm{Im}[\varepsilon_{nm}^{(0)}\delta_u\bm{v}_{nm}\times \bm{v}_{mn}+\\
  & 2v_{nm}^x v_{mn}^y (p_m-p_n)],
\end{split}    
\end{equation}
where we have recalled $\Omega_{n\bm{k}}=i\sum_{m\neq n}\bm{r}_{nm}\times \bm{r}_{mn}$. Notice that Eq.~\eqref{eq:par_mnk} and \eqref{eq:par_onk} describes how time-reversal {\it odd} quantum geometric quantities change with layer electric potential difference. Such layer potential dependent quantum geometry is possible due to tunneling between layers in $H_0(\bm{k})$: it vanishes when the layers are decoupled.

Using Eq.~\eqref{eq:par_vnm} to \eqref{eq:par_onk} and collecting terms $\mathcal{O}(EB)$ to the energy $\langle H(E,B) \rangle $, we obtain an explict form of the orbital magneto-electric moment as 
\begin{equation}
\label{eq:EBcouple}
\frac{\alpha_{n\bm{k}}}{ed}=-\delta_u m_{n\bm{k}}+2\sum_{l\neq n}\frac{\mathrm{Re}(p_{nl}m_{ln})}{\varepsilon_{nl}^{(0)}}, 
\end{equation}
where first term in Eq.~\eqref{eq:EBcouple} can be traced to changes to the Zeeman energy with respect to $E$ field. The second term in Eq.~\eqref{eq:EBcouple} arises from a $B$ field induced change to the layer dipole energy. Eq.~(\ref{eq:EBcouple}) is one of the key results of our work and reveals how electrons can intrinsically possess a magneto-electric coupling.

Interestingly, both the field induced changes to layer dipole energy and Zeeman energy contribute to variations in the layer dipole and magnetic moment. In particular, the layer dipole and orbital magnetic moment satisfy a ``microscopic'' Maxwell relation, namely
\begin{equation}
\lim_{E\to 0}\frac{\partial \langle \psi_{n\bm{k}} | \hat{\bm{p}} | \psi_{n\bm{k}} \rangle}{\partial B} = \lim_{B\to 0}\frac{\partial \langle \psi_{n\bm{k}} | \hat{\bm{m}} | \psi_{n\bm{k}} \rangle}{\partial E} = \alpha_{n {\bm k}},
\label{eq:maxwellnew}
\end{equation}
where $|\psi_{n\bm{k}}\rangle$ is the eigenstate of $H(E,B)$ and we have used the Feynmann-Hellman theorem together with Eq.~(\ref{eq:energy}). Note that the interband orbital moment $m_{nn'}$ vanishes for a two-band model zeroing the second term (dipole energy change) in Eq.~(\ref{eq:EBcouple}). Even in this limit, the microscopic Maxwell relation in Eq.~(\ref{eq:maxwellnew}) is still satisfied: counterintuitively, the magnetic field induced layer dipole depends on $E$ field induced changes to a Zeeman energy. This further demonstrates how the layer dipole is intimately intertwined with the magnetic moment.

\begin{figure}
		\centering
		\includegraphics[width=1.0\linewidth]{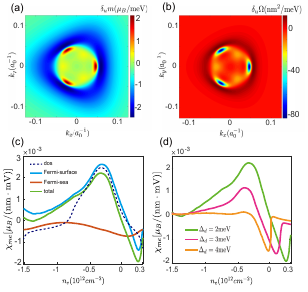}
		\caption{(a) and (b) The momentum-space $\delta_u m$ and $\delta_u \Omega$ for the hole band of RPG at $\Delta_d=4$ meV. (c) The LMC coefficient $\chi_{me}$ as a function of carrier density $n_e$ at the temperature $T=0.8$K and $\Delta_d=2$ meV. The Fermi-surface, Fermi-sea and total contributions are plotted by blue, orange and green colors, respectively. The spin degeneracy is considered in the calculation. (d) $\chi_{me}$ as a function of $n_e$ at different $\Delta_d$.}
		\label{fig:fig2}
\end{figure}

Moving to the total induced magnetization and polarization in a sample, we write 
$\delta M = - \lim_{B\to 0} \partial_B F(E,B)$ and $\delta P = - \lim_{E\to 0} \partial_E F(E,B)$ to directly obtain the linear LMC coefficients from Eq.~\eqref{eq:free_energy} as
\begin{eqnarray}
   \chi_{em} & = \chi_{me} = ed \sum_{n\bm{k}} \Big\{-\alpha_{n\bm{k}} f_{n\bm{k}}/(ed) +  m_{n\bm{k}} p_{n\bm{k}} f'_{n\bm{k}}  \nonumber \\ 
     &+ (e/\hbar) \big(\delta_u \Omega_{n\bm{k}} \phi \big[\varepsilon_{n\bm{k}}^{(0)}\big]/\beta - \Omega_{n\bm{k}} f_{n\bm{k}} p_{n\bm{k}}\big) \Big\},
   \label{eq:generalchi}
\end{eqnarray}
where $f_{n\bm{k}} = -\partial_\epsilon \phi(\epsilon)/\beta = \{1+ {\rm exp}{[\beta (\varepsilon_{n\bm{k}}^{(0)} -\mu})]\}^{-1}$ is the Fermi-Dirac function. Since the $m$ and $\Omega$ terms are time-reversal-odd, $\delta_u$ and $p$ terms are inversion-odd, both time reversal and inversion symmetries need to be broken to produce a non-vanishing $\chi_{me}$. The quantities $\delta_u\Omega_{n\bm{k}}$ and $\alpha_{n\bm{k}}$ are intrinsic, endowed by the quantum-geometric nature of Bloch electrons in the layer space. 

To be concrete, in the following we focus on the linear LMC coefficients in multilayer graphene systems, for example, as observed in RPG~\cite{han2023orbital}. The $K$ and $K'$ valleys possess opposite orbital magnetic moments and Berry curvature, as well as opposite $\delta_u \Omega_{n\bm{k}}$ and $\alpha_{n\bm{k}}$. To facilitate our analysis, we will focus on a metallic state with a valley-polarized order parameter, $\Delta_v$ ($\Delta_v \ll \mu, W$) where $W$ is the bandwidth. Such valley-polarized (metallic) magnetic order has been widely used to describe magnetic ordering in multilayer graphene via $\varepsilon_n^\tau \rightarrow \varepsilon_n^\tau + \tau \Delta_v$~\cite{hu2023josephson,huang2023spin,jang2023chirality}, where $\tau = \pm 1$ denotes the valley index. The general formula presented in Eq.~\eqref{eq:generalchi} can also be applied to study other correlated states through Hartree-Fock calculations~\cite{koh2024correlated,wang2024electrical,das2024superpolarized}.

Expanding the energies in Eq.~(\ref{eq:generalchi}), we directly find
\begin{equation}
\label{eq:lmc_coe}
\frac{\chi_{me}}{2ed\Delta_v}\!=-\hspace{-2mm}\!\sum_{n\bm{k} \in \rm{K}}\!\Big[\big(\frac{\alpha_{n\bm{k}}}{ed}
\!+\!\frac{e}{\hbar}\Omega_{n\bm{k}}p_n\big)f'_{n\bm{k}}\!+\!\frac{e}{\hbar}\delta_u \Omega_{n\bm{k}}f_{n\bm{k}}\Big].
\end{equation}
In obtaining Eq.~(\ref{eq:lmc_coe}) we have have summed over both valleys so that the momentum integral in Eq.~(\ref{eq:lmc_coe}) is only carried out within the $K$ valley; we have also used Eq.~(\ref{eq:EBcouple}) and suppressed terms going as $f''$ since it is odd in energy. Importantly, Eq.~(\ref{eq:lmc_coe}) demonstrates that the LMC in valley-polarized layered metals depends on $\Delta_v$ and have contributions that are delineated into Fermi-surface ($f'_{n\bm{k}}$) and Fermi-sea ($f_{n\bm{k}}$) contributions. 
Note that in two-band models (widely used in describing multilayer graphene) the second term in Eq.~\eqref{eq:EBcouple} vanishes. Even for multi-band model, the interband orbital moment in multilayer graphene is small~\cite{bhowal2021orbital}, making $\alpha_{n\bm{k}} \approx - ed \delta_u m_{n\bm{k}}$ a good approximation of the magneto-electric moment.

\vspace{2mm}
\emph{LMC in Rhombohedral Pentalayer Graphene.}---To demonstrate LMC
we now turn our attention to RPG as depicted by the crystal structure in Fig.~\ref{fig:fig1}(a). Close to charge neutrality, the electron wave functions in RPG are nearly localized at $A$ sites on the top layer and $B$ sites for the bottom layer. Therefore, the low-energy electronic properties of RPG can be described by the continuum Hamiltonian acting in the pseudospin basis of ($A_1,B_5$), which reads~\cite{koshino2009trigonal}
\begin{equation}
H_{\tau}(\bm{k})=\frac{v_0^5}{\gamma_1^4}\left(\begin{array}{cc}
  0 & (\pi^\dagger)^5 \\
\pi^5 & 0
\end{array}\right)+\frac{\Delta_d}{2} \sigma_z+H_R,
\label{eq:hamiltonian}
\end{equation}
where $\pi=\tau p_x+ip_y$. The inversion breaking term $\Delta_d$ is an adjustable interlayer potential difference between the top and bottom layers and $d=1.4$nm is the interlayer distance. $H_R$ is the remote hopping-term corrections. The lattice constant of graphene is $a_0=2.46\AA$. Using Eq.~\eqref{eq:EBcouple} we plot the magneto-electric moment $-\alpha_{n \bm{k}}/ed = \delta_{u}m_{n\bm{k}}$ [for the two band model in Eq.~(\ref{eq:hamiltonian})] and $\delta_{u}\Omega_{n\bm{k}}$ in the momentum space for the valence band of RPG in Fig.~\ref{fig:fig2}(a) and (b) demonstrating how the magneto-electric moment is concentrated at the band edges.

\begin{figure}
		\centering
		\includegraphics[width=1.0\linewidth]{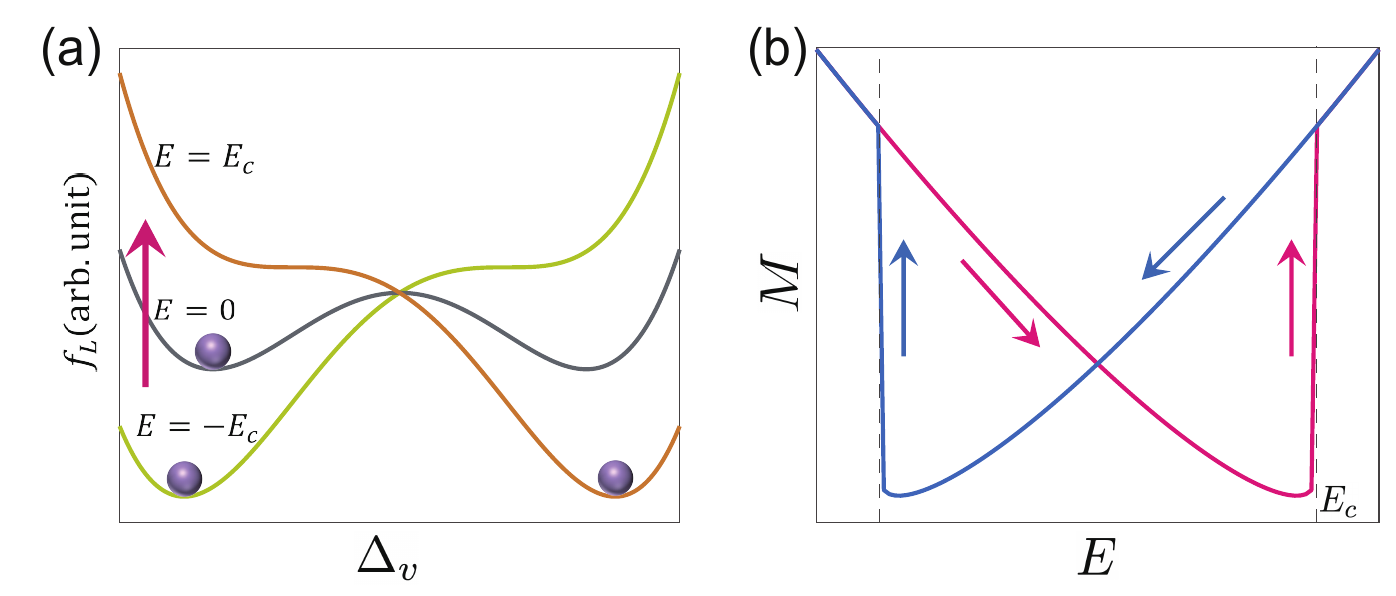}
		\caption{(a) The free energy landscapes at different electric fields. (b) The magnetization butterfly-shaped hysteresis curve induced by the electric field. The coercive field is denoted as $E_c$. (a) and (b) are evaluated phenomelogically from Eq.~\eqref{eq:landau_free}.}
		\label{fig:fig3}
\end{figure}
 
We evaluate the linear LMC coefficient $\chi_{me}$ as a function of carrier density in Fig.~\ref{fig:fig2}(c). Here we have used $\Delta_v=0.2$ meV estimated from the transition temperature for the onset of valley order in RPG~\cite{han2023orbital}. Evidently, $\chi_{me}$ is maximized near the carrier density $n_e\approx -0.4\times 10^{12}\mathrm{cm}^{-2}$ close to a peak in the density of states (dashed line). This value is close to the experimental observations ($\approx-0.5\times 10^{12}\mathrm{cm}^{-2}$)~\cite{han2023orbital}. We find that the Fermi-surface term (blue) arising from $\alpha_{n \bm{k}}$ dominates $\chi_{me}$; the Fermi-sea contribution (red) is minor. Additionally, we plot $\chi_{me}$ as a function of $\Delta_d$ in Fig.~\ref{fig:fig2}(d). Upon increasing the layer potential $\Delta_d$, $\chi_{me}$ generally decreases due to the weakened orbital magneto-electric moment. This indicates that a large LMC response is more likely to occur in the regime of small $\Delta_d$.

\vspace{2mm}
\emph{Butterfly-shaped hysteresis.}---As we now explain, LMC directly couples with valley order to produce an electric field switching of magnetic order recently seen in the experiment of Ref.~\cite{han2023orbital}. In doing so, we formulate a phenomenological Landau theory for the valley-polarized order with LMC. To understand the interplay between the electric field and valley polarization, we first adopt a small but fixed $B$ field that contributes a magneto-electric like contribution to the free-energy density: $-\delta M B-\delta P E=-(\chi_{me}+\chi_{em}) EB = -\Gamma \Delta_v E$. Such fixed B fields were present in the switching devices of Ref.~\cite{han2023orbital}. $\Gamma$ parametrizes an emergent valley-electric coupling parameter $\Gamma = 2\chi_{me}B$ mediated by the orbital LMC. As a result, the Landau free energy for the valley-polarized order can be written as
\begin{equation}
\label{eq:landau_free}
f_{L}= -a\Delta_{v}^2+\frac{b}{2}\Delta_{v}^4-\Gamma \Delta_{v}E, 
\end{equation}
where $a,b$ are phenomenological Landau parameters describing the ordering transition. In what follows, we will treat $a,b, \Gamma$ phenomenologically to demonstrate the coercive-hysteretic behavior.

\begin{figure}
		\centering
		\includegraphics[width=1.0\linewidth]{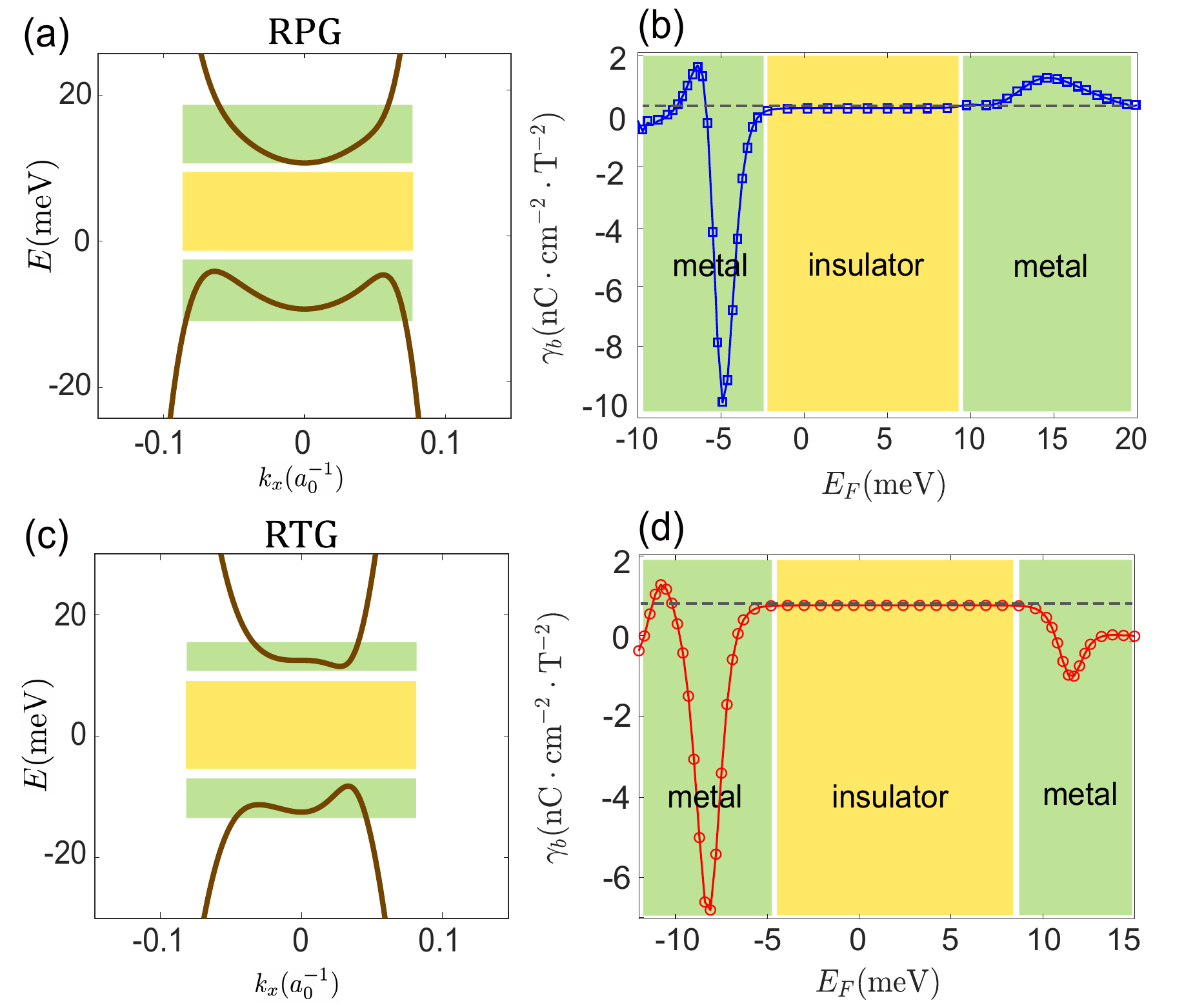}
		\caption{(a) The band structure of RPG with the layer potential $\Delta_d=20$ meV. (b) The nonlinear LMC coefficient $\gamma_b$ as a function of Fermi energy $E_F$. The insulating and metallic regions are labelled by yellow and green colors, respectively. (c) and (d) The results for RTG with the layer potential $\Delta_d=20$ meV. The temperature is set to be $T=5$K.}
		\label{fig:fig4}
\end{figure}

Below the critical temperature, we have $a>0$, $b>0$, and the valley-polarized order $\Delta_v=\pm\sqrt{a/b}$. As shown in Fig.\ref{fig:fig3}(a), the free energy landscape as a function of $\Delta_v$ evolves under different electric field strengths. Starting from a state with $\Delta_v < 0$, the order parameter evolves as an electric field is applied. When the electric field reaches coercive field ($E=E_c$) [see orange curve], the local minimum at $\Delta_v < 0$ of the free energy collapses (it becomes unstable) and transitions to a global minimum at $\Delta_v >0$. This coercive field can be written as $E_c=\frac{4a}{3\Gamma}\sqrt{\frac{a}{3b}}$. Interestingly, since $E_c$ becomes large for small $\Gamma$, the coercive fields rapidly increases for small $B$ rendering electric field induced switching challenging. This behavior is consistent with reports of increasing coercive fields with decreasing magnetic field~\cite{han2023orbital}; nevertheless modest values $B \sim 0.1 \, T$ readily produce accessible $E_c$ as seen in Ref.~\cite{han2023orbital}.

The coercive field induced transitions directly produce butterfly-shaped hysteretic curves as a function of $E$. To see this, we plot the magnetization $\delta M = \chi_{me} E$ as a function of $E$ in Fig.~\ref{fig:fig3}(b). Here magenta (blue) curves denote valley order with $\Delta_v < 0$ ($\Delta_v>0$). Because $\chi_{me}$ is locked to $\Delta_v$, the diagonal slopes of the magnetization vs $E$ curves directly reflect the valley order. In contrast, the vertical lines track the transitions between $\Delta_v < 0$ and $\Delta_v>0$ ordered states. Note the butterfly pattern flips when $B$ changes sign (i.e. $\Gamma \longleftrightarrow -\Gamma$) mirroring the flipping of the hysteretic pattern seen in Ref.~\cite{han2023orbital}. Such a butterfly has also been observed in numerical calculations~\cite{das2024superpolarized}.

\vspace{2mm}
\emph{Nonlinear magneto-electric coupling.}---While the linear orbital LMC above requires both broken time reversal and broken inversion symmetry, nonlinear LMC, on the other hand, have less stringent symmetry requirements. For example, $\gamma_e$ [Eq.~\eqref{eq:me_eq}] only requires broken time-reversal and $\gamma_b$ [Eq.~\eqref{eq:em_eq}] only requires broken inversion symmetry. As a demonstration we evaluate $\gamma_b$ for time-reversal symmetric dual-gated graphene multilayers.

Using a Green's function approach to evaluate $B$ field induced changes~\cite{fukuyama1971theory,freimuth2017geometrical} to the free energy, we obtain the nonlinear polarization (induced by $B^2$) coefficient $\gamma_b$ as
\begin{equation}
\label{eq:nonlinear_green}
\gamma_b=\frac{e^3\hbar^2}{4\beta}\sum_{n\bm{k}}\mathrm{Tr}\{\hat{\bm{p}}G[(\hat{v}_xG\hat{v}_y G)^2+(\hat{v}_yG\hat{v}_x G)^2]\},
\end{equation}
where $G=[i\omega_n+E_F-H_0(\bm{k})]^{-1}$ is the single-particle Matsubara Green’s function with $\omega_n=2(n+1)\pi/\beta$. Here $\gamma_b$ is expressed in compact Green's functions form without expanding into multiple terms~\cite{gao2015geometrical,freimuth2017geometrical} making it straightforward for numerical evaluation.

As an illustration, we evaluate $\gamma_b$ for the time-reversal symmetric phase of multilayer graphene in Fig.\ref{fig:fig4} [RPG in (b) and rhombohedral trilayer graphene in (d)]. While multilayers possess pronounced $\gamma_b$ in its metallic phase (green), most striking is the presence of a non-vanishing plateau in the insulating phase (yellow). Such a nonlinear LMC effect captures a Fermi-sea contribution and produce sizeable nonlinear polarizations. For instance, at modest magnetic field $B=2$T, we find the induced polarization $\delta P\sim 4 \mathrm{nC/cm^{-2}}$ in the insulating regime and can reach  $30\mathrm{nC/cm^{-2}}$ in the metallic regime. Interestingly, such magnetic field induced nonlinear polarizations have values comparable to nonlinear polarizations induced at second-order in the electric field ~\cite{matsyshyn2023layer,gao2020tunable,zheng2024interlayer} underscoring the strength of the LMC in multilayer graphene. 

In this work, we have presented a compact microscopic theory of LMC in multilayer graphene and elucidated its orbital (and layer degree of freedom) origin. Notably, we identify a novel orbital magneto-electric moment [Eq.~\eqref{eq:EBcouple}] as a key quantum geometric quantity (in layer space) that dominates the magneto-electric response in multilayers.  We note that the generation of nonlinear magnetization~\cite{hu2025nonlinear} and layer electric polarization has recently garnered intense recent interest~\cite{matsyshyn2023layer,gao2020tunable,zheng2024interlayer}. Our general result for nonlinear LMC highlights new opportunities for controlling layer polarization, applicable to a broad range of noncentrosymmetric layered materials~\cite{zhou2024quantum} and can be experimentally observed via capacitance probes~\cite{young2011capacitance,young2012electronic}.

\emph{Acknowledgements.}---We especially thank Cong Xiao for inspiring discussions.  This work was supported by the Ministry of Education Singapore under its Academic Research Fund Tier 2 Grant No. MOE-T2EP50222-0011 and Tier 3 Grant No. MOEMOET32023-0003 Quantum Geometric Advantage.

\clearpage
		\onecolumngrid
\begin{center}
			\textbf{\large Supplementary Material for\\ ``Orbital Longitudinal Magneto-electric Coupling in Multilayer Graphene''}\\[.2cm]		
      Jin-Xin Hu,$^{1,2}$  Justin C. W. Song$^{1}$\\[.1cm]
        {\itshape ${}^1$Division of Physics and Applied Physics, School of Physical and Mathematical Sciences, Nanyang Technological University, Singapore 637371}
         {\itshape ${}^2$Department of Physics, Hong Kong University of Science and Technology, Clear Water Bay, Hong Kong, China}

\end{center}
	
	\maketitle

\setcounter{equation}{0}
\setcounter{section}{0}
\setcounter{figure}{0}
\setcounter{table}{0}
\setcounter{page}{1}
\renewcommand{\theequation}{S\arabic{equation}}
\renewcommand{\thesection}{ \Roman{section}}

\renewcommand{\thefigure}{S\arabic{figure}}
\renewcommand{\thetable}{\arabic{table}}
\renewcommand{\tablename}{Supplementary Table}

\renewcommand{\bibnumfmt}[1]{[S#1]}
\renewcommand{\citenumfont}[1]{#1}
\makeatletter

\maketitle

\setcounter{equation}{0}
\setcounter{section}{0}
\setcounter{figure}{0}
\setcounter{table}{0}
\setcounter{page}{1}
\renewcommand{\theequation}{S\arabic{equation}}
\renewcommand{\thesection}{ \Roman{section}}

\renewcommand{\thefigure}{S\arabic{figure}}
\renewcommand{\thetable}{\arabic{table}}
\renewcommand{\tablename}{Supplementary Table}

\renewcommand{\bibnumfmt}[1]{[S#1]}
\renewcommand{\thesection}{S\arabic{section}}
\renewcommand{\theequation}{S\arabic{equation}}
\renewcommand{\thetable}{S\arabic{table}}
\renewcommand{\thefigure}{S\arabic{figure}}
\setcounter{equation}{0}
\setcounter{page}{1}

\maketitle

\makeatletter 

\section{S1: Microscopic theory of longitudinal magnetoelectric coupling}

\subsection{Layer-space quantum geometry}
For Bloch electrons in layered quantum systems, the layer-space quantum geometry gives rise to several intrinsic quantities, such as momentum-space Berry curvature $\Omega_{n\bm{k}}$, orbital magnetic moment $m_{n\bm{k}}$ and layer electric dipole moment $p_{n\bm{k}}$. Throughout this work, we focus on the orbital degree of freedom without considering spin (this is true for multilayer graphene with vanishing spin orbital coupling). In the main text, we illustrate how layer-space intrinsic quantities give rise to the longitudinal magnetoelectric coupling (LMC) coefficients. Here, we give more details to the band-summation formalism of these terms. 

Without loss of generality, we consider a layered system with the Bloch Hamiltonian $H_0(\bm{k})$. $H_0(\bm{k})$ is the bare material Hamiltonian so that $H_0(\bm{k}) |\psi_{n \bm{k}}^{(0)}\rangle = \varepsilon_{n\bm{k}}^{(0)} |n \bm{k}\rangle$ with $|n \bm{k}\rangle$ the Bloch eigenstate. Specifically, each Bloch electron for $n$th band carries Berry curvature as well as orbital magnetic moment, which read~\cite{xiao2005berry}
\begin{eqnarray}
\Omega_{n\bm{k}}&=&i\sum_{m\neq n}\bm{r}_{nm}\times \bm{r}_{mn}\\
m_{n\bm{k}}&=&\frac{e}{2}\sum_{m\neq n}\bm{v}_{nm}\times \bm{r}_{mn}
\end{eqnarray}
where the interband velocity $\bm{v}_{nm}=\langle n |\hat{\bm{v}}|m\rangle$ with $\hbar \hat{\bm{v}}=\nabla_{\bm{k}}H_0(\bm{k})$ being the velocity operator. The interband Berry connection is $r_{mn}^a=\langle m|i\partial_a|n\rangle=-i\hbar v_{mn}^a/(\varepsilon_m^{(0)}-\varepsilon_n^{(0)})$. The layer electric dipole moment is $p_{n\bm{k}}=\langle n|\hat{p}|n\rangle$ with $\hat{p}$ being the layer polarization operator. For example, $\hat{p}=1/2\sigma_z$ for a bilayer system where $\sigma_z$ is a Pauli matrix in the layer space.

Under the external vertical electric field $E$ and magnetic field $B$, the Bloch Hamiltonian changes to 
\begin{equation}
H(E,B)=H_0(\bm{k})-\hat{m} B +u \hat{p}
\end{equation}
We have used $u=eEd$ with $d$ being the thickness of the layered material. The orbital magnetic momentum operator $\hat{m} = \frac{e}{4}(\hat{v} \times \hat{r} - \hat{r}\times \hat{v})$ can be evaluated as $\langle n|\hat{m} |n'\rangle = m_{nn'}=-\frac{ie\hbar}{4}\sum_{l\neq n,n'}(1/\varepsilon_{ln}^{(0)}+1/\varepsilon_{ln'}^{(0)})\bm{v}_{nl}\times \bm{v}_{ln'}$. We can then derive the perturbation of Berry curvature and orbital magnetic moment as
\begin{equation}
\tilde{\Omega}_{n\bm{k}}=\Omega_{n\bm{k}}+u\delta_u \Omega_{n\bm{k}}, \tilde{m}_{n\bm{k}}=m_{n\bm{k}}+u\delta_u m_{n\bm{k}}
\end{equation}
Here $\Omega_{n\bm{k}},m_{n\bm{k}},\delta_u \Omega_{n\bm{k}},\delta_u m_{n\bm{k}}$  can be served as layer-space quantum geoemtrical terms of a multilayer system which can be evaluated with the bare Hamiltnian $H_0$. $\tilde{\Omega}_{n\bm{k}},\tilde{m}_{n\bm{k}}$ are the terms after the perturbation. Using the standard perturbation theory, we have $|\tilde{n}\rangle= |n\rangle +u\partial_u |n\rangle$, where $\partial_u|n\rangle = \sum_{l\neq n}|l\rangle p_{ln}/\varepsilon_{nl}^{(0)}$. For the interband velocities and Berry connections $\tilde{\bm{v}}_{nm}=\bm{v}_{nm}+u \delta_u \bm{v}_{nm}$ and $\tilde{\bm{r}}_{nm}= \bm{r}_{nm}+u \delta_u \bm{r}_{nm}$, they can be derived as 
\begin{eqnarray}
\label{eqs:par_vnm}\delta_u \bm{v}_{nm}&=&\sum_{q\neq n}p_{nq}\bm{v}_{qm}/\varepsilon_{nq}^{(0)}+\sum_{l\neq m}p_{lm}\bm{v}_{nl}/\varepsilon_{ml}^{(0)},\\
\label{eqs:par_rnm}\delta_u \bm{r}_{nm}&=&-i\hbar[\delta_u \bm{v}_{nm}\varepsilon_{nm}^{(0)}-\bm{v}_{nm}(p_n-p_m)]/[\varepsilon_{nm}^{(0)}]^2,
\end{eqnarray}
Accordingly, the perturbations of Berry curvature as well as orbital moment can be derived as 
\begin{eqnarray}
\label{eqs:par_omegac}\delta_u \Omega_{n\bm{k}}&=&i\sum_{m\neq n}(\delta_u \bm{r}_{nm}\times  \bm{r}_{mn}+\bm{r}_{nm}\times \delta_u \bm{r}_{mn})\\
\label{eqs:par_moment}\delta_u m_{n\bm{k}}&=&\frac{e}{2}\sum_{m\neq n}[\delta_u\bm{v}_{nm}\times \bm{r}_{mn}+\bm{v}_{nm}\times \delta_u\bm{r}_{mn}]
\end{eqnarray}
Then we can simplify them and obtain
\begin{equation}
\label{eqs:par_mnk}
\begin{split}
  \delta_u m_{n\bm{k}}=-\sum_{m\neq n}\frac{e\hbar}{[\varepsilon_{nm}^{(0)}]^2}\mathrm{Im}[\varepsilon_{nm}^{(0)}\delta_u\bm{v}_{nm}\times \bm{v}_{mn}+
   v_{nm}^x v_{mn}^y (p_m-p_n)],
\end{split}    
\end{equation}
and
\begin{equation}
\label{eqs:par_onk}
\begin{split}
  \delta_u \Omega_{n\bm{k}}=-\sum_{m\neq n}\frac{2\hbar^2}{[\varepsilon_{nm}^{(0)}]^3}\mathrm{Im}[\varepsilon_{nm}^{(0)}\delta_u\bm{v}_{nm}\times \bm{v}_{mn}+ 2v_{nm}^x v_{mn}^y (p_m-p_n)],
\end{split}    
\end{equation}
We have formulated a band-summation rule to evaluate the $\delta_u \Omega_{n\bm{k}}$ as well as $\delta_u m_{n\bm{k}}$. It is more readily for us to evaluate them for two-band model. Specifically, in the two-band case ($m,n=1,2$), we have $\delta_u \bm{v}_{nm}=p_{nm}(\bm{v}_{m}-\bm{v}_{n})/\varepsilon_{nm}^{(0)}$ ($\bm{v}_n=\langle n|\bm{v}|n\rangle$), and 
\begin{equation}
\label{eqs:par_connection}
\delta_u \bm{r}_{nm}=-i\hbar[\frac{p_{nm}(\bm{v}_m-\bm{v}_n)-\bm{v}_{nm}(p_n-p_m)}{[\varepsilon_{nm}^{(0)}]^2}]
\end{equation}
Inserting Eq.~\eqref{eqs:par_connection} into Eq.~\eqref{eqs:par_omegac} and ~\eqref{eqs:par_moment}, we can obtain
\begin{eqnarray}
\label{eqs:par_omega}
\delta_u \Omega_{n\bm{k}}&=& -\frac{2\hbar^2}{[\varepsilon_{nm}^{(0)}]^3}\mathrm{Im}[p_{nm}(\bm{v}_m-\bm{v}_n)\times \bm{v}_{mn}+2v_{nm}^x v_{mn}^y (p_m-p_n)]\\
\label{eqs:par_m}\delta_u m_{n\bm{k}}&=& -\frac{e\hbar}{[\varepsilon_{nm}^{(0)}]^2}\mathrm{Im}[p_{nm}(\bm{v}_m-\bm{v}_n)\times \bm{v}_{mn}+v_{nm}^x v_{mn}^y (p_m-p_n)].
\end{eqnarray}
\begin{figure}
		\centering
		\includegraphics[width=1.0\linewidth]{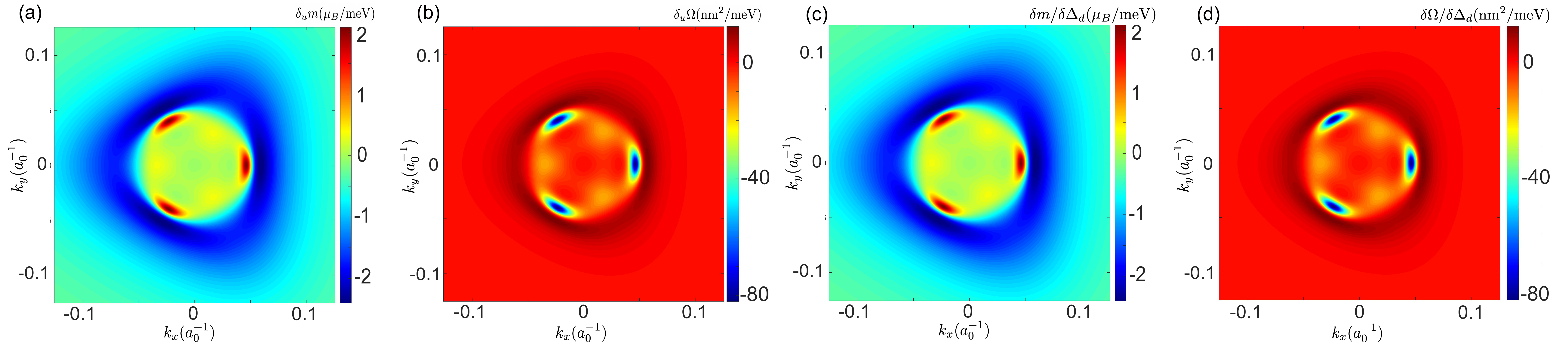}
		\caption{(a), (b) The $\delta_u m$, $\delta_u \Omega$ for pentalayer graphene (valence band) evaluated from Eq.~\eqref{eqs:par_m} and \eqref{eqs:par_omega}. (c), (d) For comparison, we calculate $[m(\Delta_d+\delta\Delta_d)-m(\Delta_d-\delta\Delta_d)]/2\delta\Delta_d$ and $[\Omega(\Delta_d+\delta\Delta_d)-\Omega(\Delta_d-\delta\Delta_d)]/2\delta \Delta_d$. }
		\label{fig:figS1}
\end{figure}
To show the validity of these formulas, we calculate the $\delta_u \Omega_{n\bm{k}}$ and $\delta_u m_{n\bm{k}}$ for the valence band of pentalayer graphene at $\Delta_d=4$ meV as shown in Fig.~\ref{fig:figS1}(a) and (b). We also directly evaluate $[m_{n\bm{k}}(\Delta_d+\delta\Delta_d)-m_{n\bm{k}}(\Delta_d-\delta\Delta_d)]/2\delta\Delta_d$ and $[\Omega_{n\bm{k}}(\Delta_d+\delta\Delta_d)-\Omega_{n\bm{k}}(\Delta_d-\delta\Delta_d)]/2\delta\Delta_d$ at $\Delta_d=4$ meV and $\delta \Delta_d=0.1$ meV in Fig.~\ref{fig:figS1}(c) and (d). Clearly, Fig.~\ref{fig:figS1}(c) and (d) are consistent with Fig.~\ref{fig:figS1}(a) and (b). As we will see, these terms are direcly related to the longitudinal magnetoelectric coupling in multilayer graphene.

On the other hand, the layer electric dipole moment also gets a correction with respect to the $B$ field. Similarly, the eigenstate is perturbed to $|\tilde{n}\rangle \rightarrow |n\rangle +B\partial_B|n\rangle$ with $\partial_B|n\rangle =\sum_{l\neq n}|l\rangle m_{ln}/\varepsilon_{nl}$. So we obtain
\begin{equation}
\tilde{p}_{n\bm{k}}=p_{n\bm{k}}+2\sum_{l\neq n}\frac{\mathrm{Re}(p_{nl}m_{ln})}{\varepsilon_{nl}}B.
\end{equation}
It is important to not that the second term arises from the interband orbital moment, which vanishes in the two-band models.

\subsection{Derivation of the LMC coefficients}
To study the magnetoelectric coupling in the layered systems, we start from the field-dependent free energy $F(E,B)$, which can be written as 
\begin{equation}
\label{eqs:free_energy}
F(E,B)=-\frac{1}{\beta}\sum_{n\bm{k}}\big\{1+\frac{e}{\hbar}B\tilde{\Omega}_{n\bm{k}}\big\}\phi\big[\varepsilon_{n\bm{k}} (E,B)\big],
\end{equation}
where $n,\bm{k}$ are the band index and wavevector, $\phi(\varepsilon)=\mathrm{ln}[1+e^{-\beta(\varepsilon-\mu)}]$. We model the field dependent Bloch band electronic energy via $\varepsilon_{n\bm{k}} (E,B)=\varepsilon_{n\bm{k}}^{(0)}+\delta \varepsilon_{n\bm{k}} (E,B)$ in an effective semiclassical fashion keeping all terms up to order $\mathcal{O}(EB)$.
Here $\delta \varepsilon_{n\bm{k}}(E,B) = -m_{n\bm{k}}B + edEp_{n\bm{k}} + \alpha_{n\bm{k}} E\cdot B$ is shown in the main text. Following the previous discussions, we can obtain $\alpha_{n\bm{k}}=(-\delta_u m_{n\bm{k}}+\partial_B p_{n\bm{k}})/ed$. Therefore, the field-induced magnetization as well as polarization can be derived as
\begin{equation}
\delta M=-\frac{\partial F(E,B)}{\partial B}|_{B=0},\delta P=-\frac{\partial F(E,B)}{\partial E}|_{E=0}.
\end{equation} 
We can then obtain the LMC coefficients as
\begin{equation}
\label{eqs:generalchi}
\chi_{me} = ed \sum_{n\bm{k}} \Big\{-\frac{\alpha_{n\bm{k}}}{ed} f_{n\bm{k}} +  m_{n\bm{k}} p_n f'_{n\bm{k}}  + (e/\hbar) \big(\delta_u \Omega_{n\bm{k}} \phi \big[\varepsilon_{n\bm{k}}^{(0)}\big]/\beta - \Omega_{n\bm{k}} f_{n\bm{k}} p_n\big) \Big\}
\end{equation}
where $f_{n\bm{k}} = \partial_\epsilon \phi(\epsilon)/\beta = \{1+ {\rm exp}{[\beta (\varepsilon_{n\bm{k}}^{(0)} -\mu})]\}^{-1}$ is the Fermi-Dirac function. It is also easy to verify the Maxwell relation $\chi_{em}=\chi_{me}$. This is a general result.

For the multilayer graphene, we need to add the valley index $\tau$. We make the assumption that, a valley-polarized order $\Delta_v$ changes the energy as $\varepsilon_{n\bm{k}}^\tau\rightarrow \varepsilon_{n\bm{k}}^\tau+\tau\Delta_v$. $\Delta_v$ can be treated as the time-reversal symmetry breaking term. We can obtain
\begin{equation}
\chi_{me}=-{2ed\Delta_v}\hspace{-2mm}\sum_{n\bm{k} \in \rm{K}}\Big[\big(\frac{\alpha_{n\bm{k}}}{ed}
+\frac{e}{\hbar}\Omega_{n\bm{k}}p_n\big)f'_{n\bm{k}}+\frac{e}{\hbar}\delta_u \Omega_{n\bm{k}}f_{n\bm{k}}\Big].
\end{equation}
In the derivation we omit the $f''_{n\bm{k}}$ term since it is odd in energy near the Fermi surface. $\delta_u m_{n\bm{k}}$ and $\delta_u \Omega_{n\bm{k}}$ have been derived in the previous subsection. 

It is worth noting that the $\chi_{me}$ should vary with respect to the $E$ field because the $E$ field changes the layer potential difference $\Delta_d$, thus the total magnetization is $M=M_0+\int_{E=0^+}^E \chi_{me}(E)dE$, where $M_0 = 0$ because $\lim_{E\rightarrow 0}\chi_{me}=0$. This was also verified in experiments~\cite{han2023orbital}. In the main text, we calculate $\chi_{me}$ at small $\Delta_d$ ($\sim$2 meV) as an estimation for the LMC at weak gate fields.

\section{S2: Phenomenological analysis of butterfy-shaped hysteresis}
In this section we provide more details of the Landau's phenomenological theory for the butterfly-shaped hysteresis observed in experiments. The valley-polarized order $\Delta_v$ can be derscribed by the standard Landau's free energy as
\begin{equation}
f_L=-a\Delta_{v}^2+\frac{b}{2}\Delta_{v}^4.
\end{equation}
where $a=a_0(T_c-T)/T_c$. Below the critical temperature $T_c$, we have $a>0$, $b>0$, and the stable solution yields $\Delta_v=\pm\sqrt{a/b}$.

We note that in the presence of applied magnetic field, the free energy gets modified as $\delta f_L=-\delta MB-\delta P E$, where $\delta M$ and $\delta P$ are electric field induced magnetization and magnetic field induced polarization, respectively. Note that because of the magnetoelectric coupling, $\delta M=\chi_{me}E$ and $\delta P=\chi_{em}B$, here we use a phenomenological way to describe the coupling term as $-\Gamma \Delta_{v}E$ as shown in the main text. Therefore, the resultant landau's energy can be written as
\begin{equation}
\label{eq:landau_free}
f_{L}= -a\Delta_{v}^2+\frac{b}{2}\Delta_{v}^4-\Gamma \Delta_{v}E.
\end{equation}
\begin{figure}
		\centering
		\includegraphics[width=0.8\linewidth]{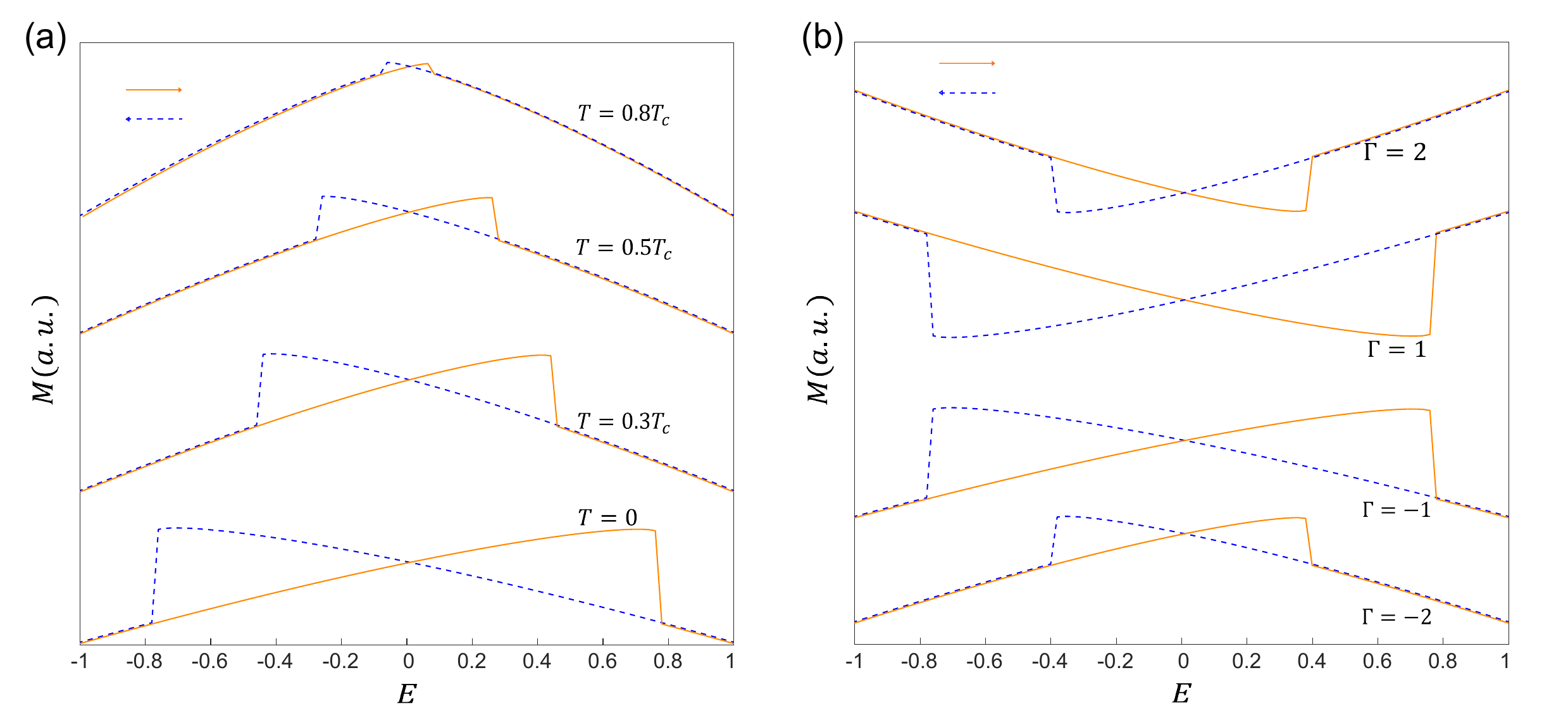}
		\caption{(a) The temperature dependence of butterfly hysteresis at $\Gamma=-1$. Near $T_c$, the butterfly gradually shrinks. (b) The magnetic field dependence of butterfly hysteresis at $T=0$. Other parameters: $a=1,b=1$.}
		\label{fig:figS2}
\end{figure}
We can solve $\Delta_v$ by $\partial f_L/\partial \Delta_v=-2a \Delta_v+2b \Delta_v^3-\Gamma E=0$. At the critical point $E_c=\frac{4a}{3\Gamma}\sqrt{\frac{a}{3b}}$, $\Delta_v$ changes its sign. By combining the LMC $\delta M \propto \Delta_v E$, we can plot the evolution of butterfly hysteresis. We examine the evolution of butterfly hysteresis under different temperature and magnetic field (equivalently, $\Gamma$), as shown in Fig.\ref{fig:figS2}. In Fig.\ref{fig:figS2}(a), we can find that as temperature $T$ increases, the coercive field $E_c$ decreases, making the butterfly area smaller. In Fig.\ref{fig:figS2}(b), the coercive field $E_c$ also decreases when $\Gamma$ is larger. Additionally, The direction of the hysteresis loop depends on the sign of $\Gamma$. These observations are consistent with the experimental measurements~\cite{han2023orbital}.

\section{S3: Nonlinear LMC in multilayer graphene}
In the presence of out-of-plane magnetic field $B$, correction of the free energy to the Bloch electrons is given by $\delta F$, which is proportional to $B^2$. $\delta F$ can be obtained as~\cite{fukuyama1971theory,freimuth2017geometrical}
\begin{equation}
\delta F=-\frac{e^2 \hbar^2 B^2}{8\beta}\sum_{p\bm{k}}\mathrm{Tr}[G(i\omega_p)\hat{v}_x G(i\omega_p)\hat{v}_y G(i\omega_p)\hat{v}_x G(i\omega_p)\hat{v}_y],
\end{equation}
where $G(i\omega_p)=[i\omega_p+\mu-H_0(\bm{k})]^{-1}$ is the Matsubara Green’s function with $\omega_p=2(p+1)\pi/\beta$. Using the definition of electric polarization, we can obtain $\delta P=-\partial \delta F/\partial E|_{E=0}$. The derivative of the Green's function is $\partial_u G=G\hat{p}G$. We note that  
\begin{equation}
\partial_u \mathrm{Tr}[G\hat{v}_x G\hat{v}_y G\hat{v}_x G\hat{v}_y]=2\mathrm{Tr}[G\hat{v}_x G\hat{v}_y \partial_u(G\hat{v}_x G\hat{v}_y)]
\end{equation}
where $\partial_u(G\hat{v}_x G\hat{v}_y)=G\hat{p}G\hat{v}_x G\hat{v}_y+G\hat{v}_x G\hat{p}G\hat{v}_y$. Therefore, the second-order LMC coefficient $\gamma_b$ can then be derived as
\begin{equation}
\label{eq:nonlinear_green}
\gamma_b=\frac{e^3 \hbar^2}{4 \beta}\sum_{p\bm{k}}\mathrm{Tr}\{\hat{p}G[(\hat{v}_x G\hat{v}_y G)^2+(\hat{v}_y G\hat{v}_x G)^2]\}.
\end{equation}
If we consider the spin and valley degeneracy of multilayer graphene, we need to multiply Eq.~\eqref{eq:nonlinear_green} by 4. In this way, the magnetic field induced electric polarization is given by $\delta P=\gamma_b B^2$.
\end{document}